\documentclass[10pt,conference]{IEEEtran}
\IEEEoverridecommandlockouts

\usepackage{amsmath,amssymb,amsfonts}

\usepackage{graphicx}

\usepackage{booktabs}

\usepackage{physics}

\usepackage{placeins}

\usepackage[font=footnotesize,labelfont=bf,justification=raggedright,singlelinecheck=false]{caption}

\usepackage[colorlinks=true,linkcolor=blue,citecolor=blue,urlcolor=blue]{hyperref}

\begin{document}

\title{Schrödinger's Toolbox: Exploring the Quantum Rowhammer Attack}

\author{\IEEEauthorblockN{Devon Campbell}
\IEEEauthorblockA{
\textit{Columbia University}\\
\textit{Department of Computer Science}\\
New York, NY, USA \\
dec2180@columbia.edu}
}

\maketitle
\begin{abstract}
   Residual cross-talk in superconducting qubit devices creates a security vulnerability for emerging quantum cloud services. We demonstrate a \emph{Clifford-only} Quantum Rowhammer attack—using just $X$ and CNOT gates—that injects faults on IBM’s 127-qubit Eagle processors without requiring pulse-level access. Experiments show that targeted hammering induces localized errors confined to the attack cycle and primarily manifests as phase noise, as confirmed by near-50\% flip rates under Hadamard-basis probing. A full-lattice sweep maps QR’s spatial and temporal behavior, revealing reproducible corruption limited to qubits within two coupling hops and rapid recovery in subsequent benign cycles. Finally, we leverage these properties to outline a prime–and–probe covert channel, demonstrating that the clear separability between hammered and benign rounds enables highly reliable signaling without error correction. These findings underscore the need for hardware-level isolation and scheduler-aware defenses as multi-tenant quantum computing becomes standard.
\end{abstract}
\vspace{-.2em}
\section{Introduction}

Quantum computing is still in the noisy intermediate-scale (NISQ) era, but real hardware is already being used for commercial and scientific tasks \cite{preskill2018quantum}. As these machines begin to handle sensitive workloads, security becomes critical \cite{portmann2022security}. Proven attack vectors now include fault injection and power-analysis side channels that expose circuit intellectual-property or corrupt results \cite{Xu2023-hd}. Because software defenses cannot block physical interference, hardware integrity is critical.

A leading hardware threat is \emph{cross-talk}—residual coupling that lets operations on one qubit disturb its neighbors \cite{abraham2013cross}. Cross-talk introduces non-local errors, degrades circuit fidelity, and worsens as devices scale, making it a major challenge for reliable multi-tenant quantum services.

Recent work shows that cross-talk can also be weaponized as an attack channel~\cite{Almaguer-Angeles2025-sq,Tan2025-sq}. Tan~\textit{et al.} demonstrated a fault-injection technique on IBM’s cloud-hosted superconducting processors that requires only standard user privileges and reliably disrupts neighboring qubits~\cite{Tan2025-sq}. However, their method depends on Qiskit Pulse, a low-level control interface that IBM is deprecating~\cite{ibm2025pulse}, partly due to such security risks.

Importantly, low-level access is not essential. High-level circuits composed of native gates can induce faults via repeated cross-talk, closely paralleling the classical \emph{rowhammer} attack on DRAM~\cite{Kim2024-ib,Mutlu2023-wt}. In that setting, rapid access to specific memory rows leaks charge into adjacent rows, flipping bits without direct access to the target data.

Analogously, densely driving “control” qubits with standard gates can leak errors into neighboring qubits, degrading computation fidelity. Almaguer-Ángeles~\emph{et al.} showed that this “Quantum Rowhammer” attack succeeds using only Clifford gates, without pulse-level commands~\cite{Almaguer-Angeles2025-sq}. Unlike Qiskit Pulse, Clifford gates are fundamental for quantum computation and must be exposed to users. Thus, these cloud systems remain inherently vulnerable to cross-talk-based fault injection that exploits the native gate set itself.

In this work, we confirm that a minimal Clifford gate set is sufficient to conduct the Quantum Rowhammer (QR) attack. No pulse-level controls are needed: a carefully timed burst of CNOTs on aggressor qubits reliably induces bit-flip errors in adjacent victims, reproducing the fidelity loss observed in the original study.

To characterize the type of errors introduced by cross-talk, we repeated the attack in the $X$-basis by applying Hadamard gates before and after measurement. The resulting flip rates, tightly clustered around 50\%, indicate that the relative phase was randomized by the QR-induced disturbance, leaving both outcomes equally likely—a signature consistent with dephasing-dominated noise.

This work makes the following key contributions:
\begin{enumerate}
    \item \textbf{Gate-level Quantum Rowhammer on modern hardware.} We demonstrate that a minimal Clifford gate set (\texttt{X}, CNOT) suffices to mount a Quantum Rowhammer (QR) attack on IBM's 127-qubit Eagle processors, without requiring deprecated pulse-level access.
    
    \item \textbf{Error characterization in the $X$-basis.} By repeating the attack with Hadamard pre- and post-rotations, we show that induced faults cluster near a 50\% flip probability, consistent with phase-randomization and dephasing-dominated noise.
    
    \item \textbf{Full-lattice spatiotemporal mapping.} We provide the first lattice-wide sweep quantifying QR's spatial locality (within two coupling hops) and temporal confinement (single-cycle recovery), across 40 alternating hammer/benign rounds.
    
    \item \textbf{Covert-channel feasibility.} We estimate a low-bandwidth cross-talk covert channel based on center-qubit modulation, achieving up to $\sim$460~bps under realistic round timings, and discuss encoding strategies.
    
\end{enumerate}

\section{Background}
\begin{figure*}[t]
  \centering
  \includegraphics[width=0.8\textwidth]{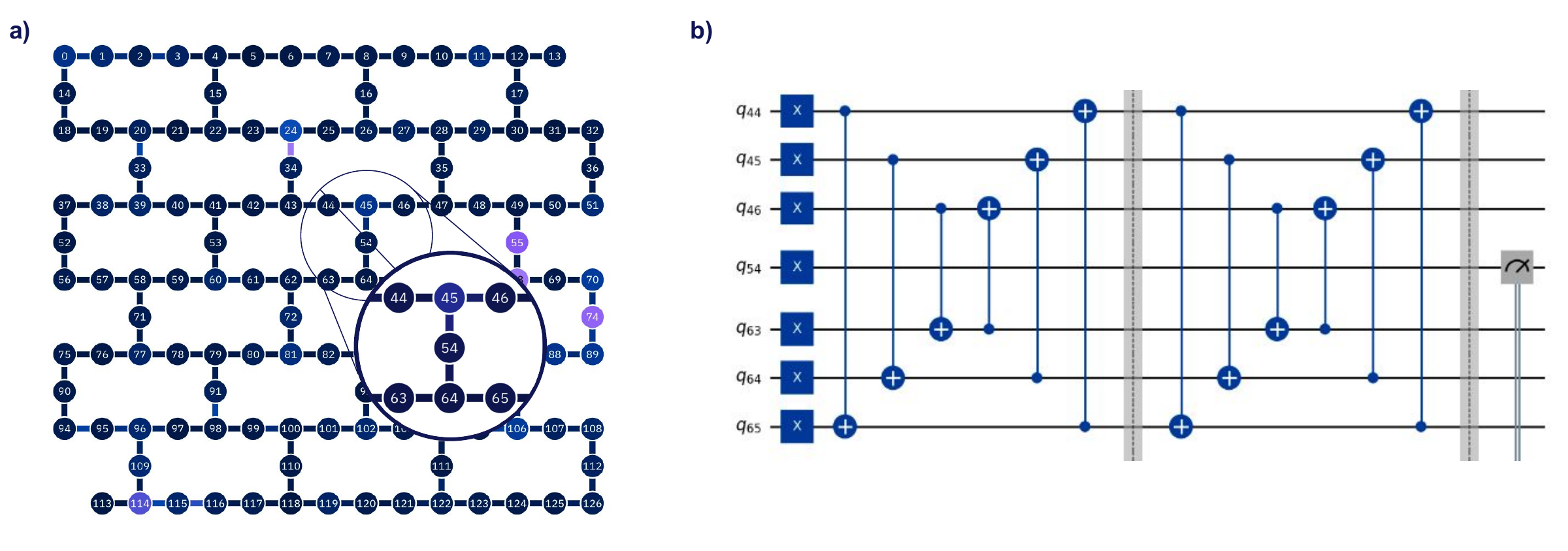}
  \caption{
    \textbf{Quantum Rowhammer Circuit.}
    \textbf{(a)} Hardware coupling map of IBM’s 127-qubit Eagle device. The “neighborhood” for qubit 54 is enlarged, illustrating the control-target layout for the experiments.
    \textbf{(b)} Example circuit schematic for the attack on qubit 54.  }
\vspace{-1em}
  \label{fig:qiskit}
\end{figure*}

Noisy Intermediate-Scale Quantum (NISQ) devices typically feature 10–100 qubits and exhibit high error rates from decoherence, gate infidelity, readout noise, and cross-talk~\cite{preskill2018quantum}. Despite these challenges, they can perform tasks that are difficult for classical computers. For example, IBM's 127-qubit Eagle processor (Fig.~\ref{fig:qiskit}) uses a heavy-hexagonal lattice of superconducting transmon qubits, where each qubit connects to one to three neighbors~\cite{Javadi-Abhari2024-ej}.

Major providers, including IBM, Amazon, and Microsoft, deliver Quantum-as-a-Service (QaaS), granting remote access via APIs and SDKs. As demand grows, vendors explore \emph{multi-tenant} scheduling, where circuits from different users share a single processor either sequentially or simultaneously on separate qubit subsets~\cite{Das2019-gt}. Multi-tenancy is not yet standard but is expected as hardware scales. While this development promises better utilization, it also introduces security concerns: cross-talk ignores logical job boundaries, enabling an adversarial workload to disrupt neighboring circuits.

\section{Threat Model}

This work considers a threat model motivated by trends in quantum cloud computing, especially the move toward scalable, shared hardware. While today’s superconducting processors (e.g., IBM’s 127-qubit heavy-hex device in Fig.~\ref{fig:qiskit}) do not yet support true multi-tenant execution—where circuits from multiple users run concurrently on shared hardware—this model is expected as demand grows~\cite{Das2019-gt}.

Here, we examine the security risks of co-located circuit execution, drawing parallels to classical cloud co-residency attacks. The adversary is an ordinary cloud user with no privileged access or physical control, limited to submitting quantum circuits built from the provider’s native gate set (e.g., Clifford operations and standard single- and two-qubit gates). The victim is another tenant running circuits on adjacent qubits at overlapping times.

The attacker’s goal is to degrade the victim’s computational fidelity while submitting a valid, policy-compliant circuit. The attack exploits physical cross-talk—unintended residual coupling between neighboring qubits—to inject faults indirectly. By applying dense sequences of standard gates (e.g., $X$, CNOT) to “aggressor” qubits, the attacker induces decoherence and stochastic errors on adjacent “victim” qubits.

\subsection{Attack Scenarios}

\paragraph{Co-Resident Sabotage}  
A victim and attacker job are scheduled simultaneously on adjacent qubits. The attacker submits an innocuous-looking circuit with an intense burst of CNOTs on its qubits—the \emph{single-center QR} experiment. Cross-talk during the victim’s computation flips its qubit states, distorting results without leaving classical traces.

\paragraph{Wide-Sweep Denial of Coherence}  
A determined adversary can submit multiple QR jobs, each targeting different control qubits, effectively sweeping the fault-inducing region across the chip. This mirrors classical rowhammer amplification techniques—such as Sledgehammer—which induce widespread corruption by hammering multiple regions simultaneously, escalating the impact beyond isolated bit-flips~\cite{yaeglici2022sledgehammer}.

\section{Methodology}

\label{sec:methodology}

This section describes the framework used to (i) launch the QR attack, (ii) sweep it across the heavy‑hex lattice, and (iii) quantify its spatial and temporal behavior. 

All experiments were executed on cloud-accessible superconducting backends (IBM Brisbane and Sherbrooke) with 127 physical qubits and the provider’s default transpiler; current schedulers do not yet support concurrent multi-tenant execution, but this is expected as hardware scales. For any chosen center qubit, its six nearest neighbors in the heavy-hex coupling graph (two direct bus connections plus their immediate upstream and downstream partners) define the \emph{hammer neighborhood} (see Fig.~\ref{fig:qiskit}a).

\subsection{Attack Protocols}

\paragraph*{Single-Target}

Each attack centers on a qubit left idle while its neighbors undergo gate-dense circuits. These center qubits were selected for their high connectivity, positioned along vertical connector links for maximal proximity to adjacent qubits (see Fig.~\ref{fig:qiskit}a). The hammer protocol proceeds in five stages (Fig.~\ref{fig:qiskit}b):

\begin{enumerate}
    \item \textbf{Initialization:} Prepare the center qubit in $\ket{0}$ or $\ket{1}$ (or, for the $X$-basis study, in $\ket{+}$ or $\ket{-}$ via Hadamard).
    \item \textbf{Neighbor burst:} Apply an $X$-gate to each of the six neighbors.
    \item \textbf{Pairwise entangling rounds:} Entangle neighbors with CNOT gates in symmetric pairs, repeated for a fixed number of rounds to focus residual $ZZ$ interactions on the idle center.
    \item \textbf{Basis undo (optional):} Apply a second Hadamard to return from $\ket{\pm}$ to the computational basis if needed.
    \item \textbf{Measurement:} Measure the center qubit and record the logical flip rate over 20,000–40,000 shots.
\end{enumerate}

\paragraph*{Lattice-Wide Sweep}

To study spatial propagation, we extended the hammer protocol to multiple, disjoint center sets. Each cycle hammered a new group of centers in parallel while all qubits on the chip were measured. Hammer and benign cycles were interleaved as internal controls to assess whether induced errors persisted over time.

\section{Results}

\subsection{Single-Target Quantum Rowhammer}
\label{subsec:qr}

\begin{figure*}[ht]
\centering
  \includegraphics[width=.85\linewidth]{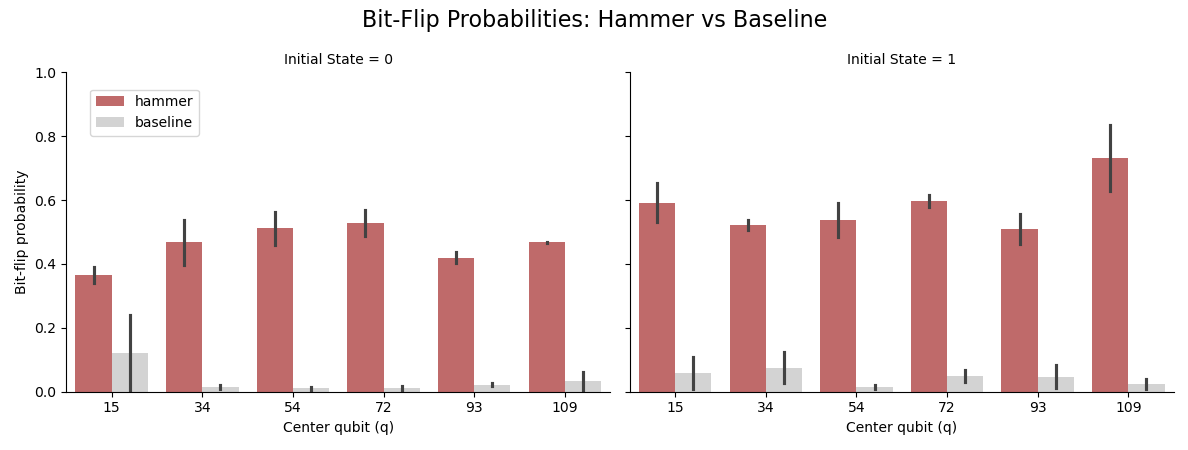}
  \caption{Bit-flip probabilities for target qubits under baseline conditions and Quantum Rowhammer attack. Each subplot shows results for $\ket{0}$ (left) and $\ket{1}$ (right). Red bars indicate flip probability with hammering; gray bars show the baseline without hammering.} 
  \vspace{-1em}
  \label{fig:hammer_vs_baseline}
\end{figure*}

To evaluate vulnerability to QR attacks, we conducted targeted fault-injection experiments. For each trial, a \emph{center} qubit was initialized in either $\ket{0}$ or $\ket{1}$ and subjected to a QR circuit with aggressive neighbor gating. A baseline circuit without hammering was also run for comparison. The flip probability—the chance that measurement yields the opposite of the initial state—was recorded.

Figure~\ref{fig:hammer_vs_baseline} reports bit-flip probabilities for selected center qubits. In all cases, hammering substantially increased error rates compared to baseline. For initial state $\ket{0}$ (left), flip probabilities consistently exceeded 40\%, often surpassing 50\%, while baseline rates stayed below 15\% (often under 5\%).

For initial state $\ket{1}$ (right), the effect was similar but showed slightly more variability across qubits. Notably, center qubit 109 exceeded 70\% flip probability under attack, highlighting location-dependent cross-talk effects.

Overall, qubits initialized in $\ket{1}$ showed higher average flip rates ($\approx 0.58$) than those in $\ket{0}$ ($\approx 0.46$). This aligns with the fact that superconducting qubits experience more spontaneous relaxation ($T_1$ decay) from $\ket{1}$ to $\ket{0}$ than excitation in the reverse direction, making $\ket{1}$ states more susceptible to cross-talk–induced errors.

\subsection{Hadamard Basis}
\label{subsec:h_basis}

\begin{figure}[ht]
\centering
  \includegraphics[width=.95\linewidth]{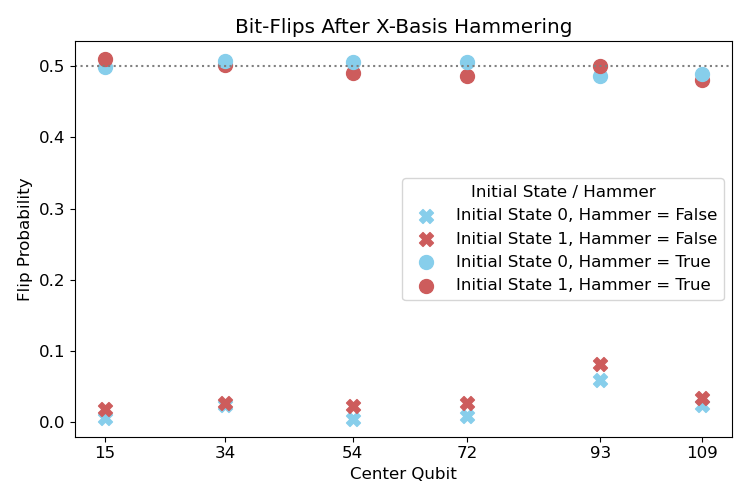}
  \caption{%
\textbf{Bit-flip rates after Hadamard hammering.}
Each point shows a center qubit’s flip probability. Crosses are controls without hammering; dots include hammer bursts. Hammering raises flip probability to $\approx50\%$, while controls stay below 3\%.
}
\vspace{-1.5em}
  \label{fig:x_bf}
\end{figure}

Figure~\ref{fig:x_bf} shows the effect of QR in the $X$-basis, where each target qubit is rotated via a Hadamard gate before and after the hammer sequence. Under hammering, flip probabilities cluster tightly around 50\% (${p_{\text{flip}}} = 0.498$, $\sigma < 0.013$) regardless of qubit index or initial state, indicating strong phase randomization consistent with cross-talk–induced dephasing. In contrast, control runs without hammering yield negligible flips (${p_{\text{flip}}} = 0.027$), confirming the attack’s effect.

Table~\ref{tab:compare-bases} compares these results with those in the $Z$ basis. While $Z$-basis flip rates vary widely depending on qubit and initial state (from $0.34$ to $0.84$), $X$-basis flip rates remain tightly centered near $0.50$. This contrast supports the hypothesis that QR introduces \emph{phase errors}, which become deterministic bit-flips in the conjugate basis.

\begin{table}[h]
\centering
\caption{Average bit-flip probabilities under hammering in the computational (Z) and Hadamard-shifted (X) bases, compared with prior work (\textit{Almaguer-Angeles et al.}). Larger ${p_{\text{flip}}}$ values correspond to stronger attack impact.}% ~\cite{Almaguer-Angeles2025-sq}.}
\label{tab:compare-bases}
\begin{tabular}{llcc}
\toprule
Basis & Initial state & ${p_{\text{flip}}}$ (This work) & ${p_{\text{flip}}}$ (Prior work) \\
\midrule
Z & $0$     & 0.455 $\pm$ 0.080 & 0.271 $\pm$ 0.232 \\
Z & $1$     & 0.592 $\pm$ 0.130 & 0.387 $\pm$ 0.255 \\
X & $0$/$1$ & 0.498 $\pm$ 0.013 & --- \\
\bottomrule
\end{tabular}
\end{table}

Qubits in the $X$-basis are sensitive to phase flips (e.g., $\ket{+} \xrightarrow{\text{Z error}} \ket{-}$). Frequent phase disturbances such as $Z$-errors and $ZZ$-interactions effectively randomize a qubit’s phase in this basis. When rotated back to the $Z$-basis via Hadamard, this phase randomization appears as an equal chance of measuring $\ket{0}$ or $\ket{1}$. The consistency across qubits and states suggests that the QR attack primarily induces dephasing.

\subsection{Rowhammer Sweep Across the Lattice}
\begin{figure*}
\centering
  \includegraphics[width=\linewidth]{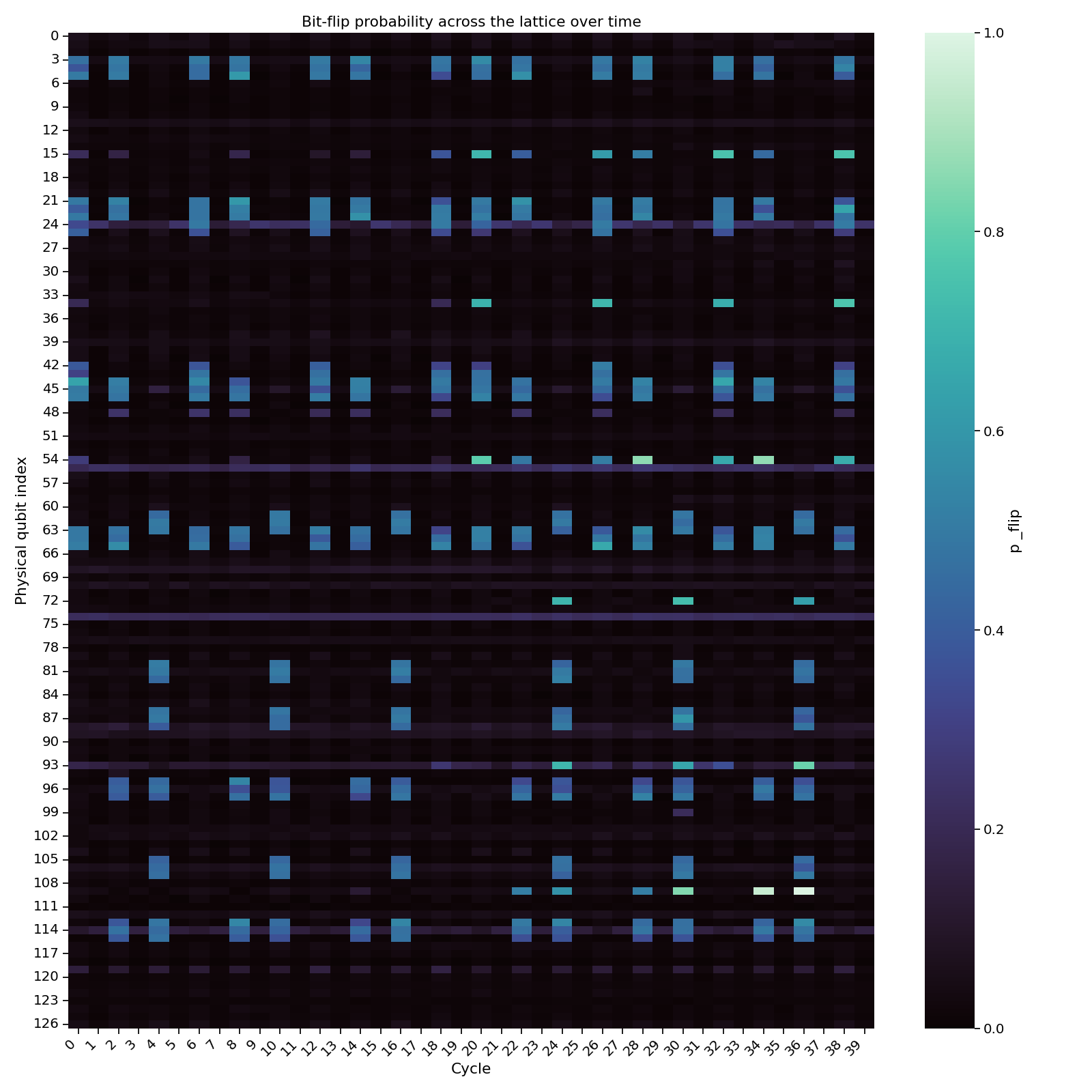}
  \caption{%
    \textbf{Lateral and temporal spread of Quantum Rowhammer.}
    The heat-map shows the measured bit-flip probability $p_{\text{flip}}$
    for every physical qubit ($y$-axis) over 40 execution cycles ($x$-axis) on
    \texttt{ibm\_brisbane}.  
    Even-numbered cycles contain a multi-center hammer burst; odd cycles
    run a benign circuit.
    Bright pixels reveal qubits whose fidelity is strongly degraded by the
    attack; dark pixels denote baseline error rates below 0.05.
    High-error regions appear only during hammer cycles and remain confined to
    qubits that are at most two hops from the driven centers, while the device
    recovers to nominal behavior in the immediately following benign round.
  }
  \label{fig:lattice-sweep}
\end{figure*}

\begin{table}
\centering
\caption{Cycle-level maximum $p_{flip}$ statistics for hammered vs benign conditions.}
\label{tab:pflip-stats}
\begin{tabular}{lrrr}
\toprule
Condition & Min $p_{flip}$ & Max $p_{flip}$ & Mean $p_{flip}$ \\
\midrule
Hammered & 0.500 & 0.993 & 0.674 \\
Benign   & 0.204 & 0.261 & 0.237 \\
\bottomrule
\end{tabular}
\end{table}

Figure~\ref{fig:lattice-sweep} shows bit-flip probabilities for all 127 qubits over forty alternating cycles of aggressive multi-center hammering and benign reference runs. Several trends emerge.

First, corruption is local, not global. During hammer cycles (even-numbered columns), qubits within or directly coupled to the targeted center group show flip probabilities up to $p_{\text{flip}} \approx 0.8$ (bright teal). The disturbance typically extends one or two coupling-graph hops beyond the driven region but does not spread across the entire lattice.

Second, recovery is rapid. In nearly all cases, the subsequent benign cycle (odd-numbered columns) returns affected qubits to the background error floor ($p_{\text{flip}} < 0.05$, black). This indicates that the cross-talk-induced errors impair victim computations only during the attack window.

Third, the attack is reproducible. The spatial envelope of elevated error closely tracks the scheduled center group across cycles, showing deterministic and repeatable fault injection rather than random device drift.

The separation between hammered and benign conditions can also be quantified at the cycle level. Table~\ref{tab:pflip-stats} reports the maximum observed $p_{\text{flip}}$ per cycle for both cases. Hammered cycles range from 0.50 to 0.99 (mean 0.67), whereas benign cycles remain tightly clustered between 0.20 and 0.26 (mean 0.24). This gap enables a simple threshold-based classifier at $p_{\text{flip}}\approx 0.30$ to perfectly separate the data. Consequently, a covert channel based on hammer presence or neighborhood selection can transmit data without error-correcting codes and maintain high reliability even under moderate background noise increases.

These findings suggest that Quantum Rowhammer can disrupt co-resident workloads in time-sliced, cloud-based settings, but its impact is both localized and short-lived, confined to the attacker’s execution window. Critically, this strong temporal and spatial separation also produces a clear statistical gap between hammered and benign cycles, as shown in Table~\ref{tab:pflip-stats}, laying the groundwork for covert signaling.

\section{Discussion}

\paragraph*{Physical Origin and Spatial Footprint}
The data show that bursts of standard Clifford operations can deterministically inject errors into neighboring qubits. Figure~\ref{fig:lattice-sweep} reveals elevated flip probabilities ($p_{\text{flip}}\geq 0.35$) appearing only during even-numbered hammer cycles, forming clear vertical bands tied to driven center qubits.

These high-error regions align with the device’s heavy-hex coupling topology, typically extending one or two hops from targeted sites. This spatial pattern suggests residual $ZZ$-type interactions along shared bus resonators as the primary mechanism for cross-talk \cite{AlmaguerAngeles2022-crosstalk}.

Importantly, the effect is both localized and transient. Nearly all affected qubits return to baseline error rates in the following benign cycle. This rapid recovery indicates the noise stems from controllable, temporary phase randomization—not permanent calibration drift or thermal damage.

Such precise temporal control yields a nuanced attack profile: the faults are spatially targeted and time-bounded, making them hard to detect and highly repeatable. This subtlety enhances the attack’s stealth, though its short-lived nature constrains its impact to specific execution windows.

\paragraph*{Effectiveness of Compiler-Level Defenses}
Implementing DD and repetition-code post-selection shows measurable error reductions, but neither approach fully neutralizes cross-talk-induced faults. DD sequences mitigate some decoherence, but their effectiveness depends on specific qubit interactions and noise environments \cite{Niu2024-vl}. Additionally, both DD and post-selection introduce trade-offs like increased execution time and reduced throughput, which may not be acceptable in all contexts. Given these limitations, software-level defenses alone are insufficient. Hardware-based solutions to improve qubit isolation may be required for robust protection \cite{Tan2025-sq}.

\paragraph*{Implications for Multi-Tenant Processors}
While current cloud schedulers time-slice jobs, larger devices in the near future are expected to host mutually unaware tenants on disjoint sub-lattices simultaneously. These results show spatial isolation is porous: a malicious tenant can degrade a neighbor’s logical fidelity by up to $\approx 80\%$. More subtly, an attack need not cause total failure—\emph{predictable} noise can bias expectation values in variational algorithms or undermine error-mitigated computations. This vulnerability underscores the need for stronger isolation protocols and security measures in shared quantum computing environments \cite{Lee2025-tq}.

\subsection{Prime–and–Probe Covert Channel Feasibility}
\label{subsec:covert-channel}

The lattice-sweep experiment (Fig.~\ref{fig:lattice-sweep}) demonstrates two properties essential for covert signaling: (i) \textbf{temporal confinement}, where cross-talk-induced errors vanish immediately after the hammer cycle, and (ii) \textbf{spatial locality}, where only qubits within $\leq 2$ hops of the driven centers exhibit elevated flip rates. Combined with the clear separation between hammered and benign cycles reported in Table~\ref{tab:pflip-stats} (hammered: 0.50–0.99, benign: 0.20–0.26), this makes threshold-based detection trivial, ensuring reliable signaling without error correction.

\paragraph*{Channel model}
Let each \emph{hammer–probe} pair form one signaling round. The sender (\emph{attacker}) primes the lattice by hammering a chosen set of centers during the hammer phase; the receiver (\emph{victim}) probes in the next cycle by running a measurement-only circuit across its allocated sub-lattice. Elevated error in the receiver’s neighborhood indicates a logical `1`; absence indicates `0` (\emph{on–off keying}). This binary modulation can be extended to multi-ary encoding by varying \emph{which} neighborhood is hammered.

\paragraph*{Encoding capacity}
Assume $m$ disjoint neighborhoods that can be targeted without overlapping cross-talk. Each round can encode
\(
R = 1 + \log_2 m 
\)
bits per round where the first bit comes from presence/absence and the rest from index selection. For $m=12$, $R \approx 4.6$~bits/round. At modest cycle times (e.g., 10~$\mu$s/round), this allows a covert bandwidth of $\sim$460 bits per second—enough to exfiltrate cryptographic keys, variational parameters, or calibration secrets across tenant boundaries.

\section{Conclusion}

This work shows that simple, Clifford-only circuits can mount a Quantum Rowhammer (QR) attack on modern superconducting qubit hardware. The induced faults are primarily phase errors, appearing as near-maximal bit-flip rates in the conjugate basis. These errors are spatially localized—within one or two hops of the attacker—and temporally precise, emerging only during targeted hammer cycles. This combination of locality, repeatability, and stealth makes QR viable both as a sabotage tool and a low-bandwidth covert channel.

Effective protection will likely require a layered strategy spanning hardware, middleware, and scheduling policies. On the hardware side, IBM's deployment of tunable bus couplers can dynamically disable qubit links and suppress residual $ZZ$-interactions. Their first production implementation in the 133-/156-qubit Heron family reportedly achieves “practically zero cross-talk’’ \cite{IBMHeronBlog2024}. If robust, this represents a major step toward closing the exploited side channel.

This study demonstrated the potential for a cross-talk-based channel in a controlled setting but did not implement a full sender/receiver protocol. A natural next step is a proof-of-concept “Prime-and-Probe” attack on today’s cloud back-ends, where a sender primes selected centers and a receiver probes afterward to decode the spatial error pattern. Such work would help quantify real-world bandwidth, reliability, and stealth.

Ultimately, Quantum Rowhammer highlights a critical lesson for secure quantum computing: as access to shared hardware grows, so does the risk of physical-layer interference between users. Even standard gate primitives, used adversarially, can degrade neighboring workloads or leak information across logical boundaries. As quantum devices scale and scheduling becomes increasingly parallel, defending against such side channels will be as essential as improving qubit fidelity or algorithmic depth. These findings underscore the need for quantum-aware security engineering at every level of the stack.

\bibliographystyle{IEEEtran}
\bibliography{bib}

\end{document}